# Polarimetric Light-Pulse Atom Interferometer: Two-Level Scheme

**A. Zh. Muradyan***

*Yerevan State University, Yerevan, Armenia*
*\*e-mail: muradyan@ysu.am*

**Abstract**—A new type of atomic interferometer is proposed, in which the traditional method of measuring the state of an atom is replaced by the technique of polarization spectroscopy using the working substance of a clot of condensate of two-level atoms. As a result, the atomic interferometer is freed from needing the above-mentioned multiple repetitions, while maintaining high sensitivity. The Kapitsa-Dirac resonance diffraction is used to split the translational motion of the atom. Numerical computations to determine the rotated component of the probing field show that the ratio of the output signal to the input signal under normal conditions in a specialized laser physics laboratory using a clot of atomic condensate of alkali metals with a concentration of $10^{11}$ cm$^{-3}$ and linear dimensions of the order of 1 μm as the working substance reaches quite respectable values.

**Keywords:** light-pulsed atomic interferometer, polarization spectroscopy, atomic condensate, Kapitsa−Dirac diffraction

## 1. INTRODUCTION

Light-pulse atomic interferometers [1–6] are an established high-tech field in quantum metrology and quantum sensing [7–9]. In them, the atomic wave packet is split, as in standard optical interferometers, into two paths, which are then reflected and recombined. The splitting angle is formed by Raman scattering of a photon on an atom, as a result of which the split atomic trajectories are very close. To move them apart, it is necessary to carefully accumulate a multitude of scattering events, up to several hundred [10, 11].

Another qualitatively new aspect of the traditional atomic interferometer about traditional optical interferometers is the method of registration. Here, information about the phase difference is determined by counting the atoms registered at their ground and excited internal energy levels (output ports), directly appealing to the standard theory of quantum measurement [12–14]. The probabilistic nature of the latter results in the need for multiple repetitions of the entire interaction cycle, including identical initial conditions.

In this paper, a new scheme of an atomic interferometer is presented, in which information about the output state of an atom is determined not by counting atoms at two output ports of the interferometer, but by a single act of highly sensitive polarization spectroscopy of the probing field [15, 16]. Another feature of the proposed scheme is that the interference of atomic waves occurs not because of the spatial overlap of two previously separated trajectories, but due to optical transitions between two discrete families of pulsed states of the translational motion of an atom in its ground and excited internal states, generated by Kapitsa-Dirac diffraction. A similar scheme of a light-pulse atomic interferometer, in which the registration of interference information is carried out by the method of absorption spectroscopy of probing radiation, was published by us [17]. The motivation for this study is the significantly higher sensitivity of polarization spectroscopy compared to absorption spectroscopy.

At the first stage of the interferometer operation, a family of equidistant pulse states is generated by the diffraction of the atom on the resonant field of the standing wave of laser radiation. To compute the amplitudes of these states, the well-known Raman−Nath approximation is extended to longer interaction times, which allows the ground and excited internal states of the atom to be populated almost equally. In the second stage, the interaction occurs with a traveling wave, generating interference between families of momentum states. During this interaction, the momentum distributions at the internal levels periodically oscillate, displacing the distribution centers in opposite directions. Following the pumping field, the prob-ing field of ultrashort coherent radiation propagates along the direction of standing and traveling waves and, interacting with the atomic ensemble, projects its quantum state into the spectrum of rotation of the plane of polarization of the probing field. The required information about the interference of matter waves is presented in the form of an asymmetrically distributed family of narrow maxima, equally shifted relative to each other because of the Doppler effect. The use of such a diagnostic method will significantly simplify the design of the atomic interferometer while maintaining high measurement accuracy.

## 2. GENERATION AND INTERFERENCE OF PULSED STATES OF AN ATOM

### 2.1. Generation of Pulsed States of an Atom by a Standing Wave Field

Let us consider a two-level atom with mass $M$ and the optical transition frequency $\omega_0$, interacting with a laser standing wave with a resonant frequency $\omega = \omega_0$ and electric field strength $E$. Note that throughout the article the time of interaction of the atom with external fields is considered to be less than the relaxation time of the atomic state. Then the state of the atom is described by the wave function

$$\psi(z,t) = g(z,t)\varphi_g e^{-iE_g t} + e(z,t)\varphi_e e^{-iE_e t},$$

where $\varphi_g$ and $\varphi_e$ is the wave functions of internal states, and $e(z,t)$ is the amplitudes (coefficient wave functions of translational motion) of an atom located respectively at the ground and excited internal energy levels $E_g$ and $E_e$ at a point in time $t$ and with the coordinate of the center of gravity $z$.

The evolution equations of these amplitudes are well known and can be written as equations separately for superposition amplitudes $a(z,t) = (g(z,t) + e(z,t))/2$ and $b(z,t) = (g(z,t) - e(z,t))/2$:

$$\left(i\frac{\partial}{\partial t} + \frac{\hbar}{2M}\frac{\partial^2}{\partial z^2}\right)a(z,t) = -2\frac{dE}{\hbar}\cos kz\, a(z,t), \tag{1a}$$

$$\left(i\frac{\partial}{\partial t} + \frac{\hbar}{2M}\frac{\partial^2}{\partial z^2}\right)b(z,t) = 2\frac{dE}{\hbar}\cos kz\, b(z,t), \tag{1b}$$

where $d$ is the dipole matrix element of optical transition, $k = \omega/c$.

First, let us consider equation (1a) and introduce dimensionless variables $\eta = kz$, $\tau = \omega_r t$ and the intensity parameter of waves $\varsigma = dE/\hbar\omega_r$, where $\omega_r = \hbar k^2/2M$ is the recoil frequency of single-photon absorption (or emission). Let us assume that before the interaction with the standing wave, the atom was in a state of rest or a superposition of momentum states differing from each other by a value multiple of the photon momentum. Then we can seek the solution of (1a) and (2b) in the form of Fourier expansions

$$a(\eta,\tau) = \sum_{n,m=-\infty}^{\infty} i^m a_m(n,\tau) e^{i(m+n)\eta - i(m+n)^2\tau}, \tag{2a}$$

$$b(\eta,\tau) = \sum_{n,m=-\infty}^{\infty} i^m b_m(n,\tau) e^{i(m+n)\eta - i(m+n)^2\tau}, \tag{2b}$$

where $|m|$ is the recoil impulse (in $\hbar k$ units), obtained by the atom in the process of interaction with the standing wave, and the possible values $n$ ($n$ not necessarily an integer) differ by one. It is used to represent the initial momentum distribution of the atom.

Substituting (2a), (2b) into (1a), and (1b), for unknown amplitudes $a_m(n,\tau)$, $b_m(n,\tau)$ we obtain the following recurrent differential equation:

$$\frac{\partial a_m(n,\tau)}{\partial \tau} = e^{-i\tau}\varsigma\left(a_{m-1}(n,\tau)e^{i2(m+n)\tau} - a_{m+1}(n,\tau)e^{-i2(m+n)\tau}\right),$$

$$\frac{\partial b_m(n,\tau)}{\partial \tau} = -e^{-i\tau}\varsigma\left(b_{m-1}(n,\tau)e^{i2(m+n)\tau} - b_{m+1}(n,\tau)e^{-i2(m+n)\tau}\right).$$

Because of the time-dependent exponential coefficients, the equations do not have exact analytical solutions. The well-known Raman–Nath approximation [18–21], sometimes called the short interaction time approximation, corresponds to replacing the coefficients by unity, i.e., when $(2m + 2n \pm 1)\tau \ll 1$.

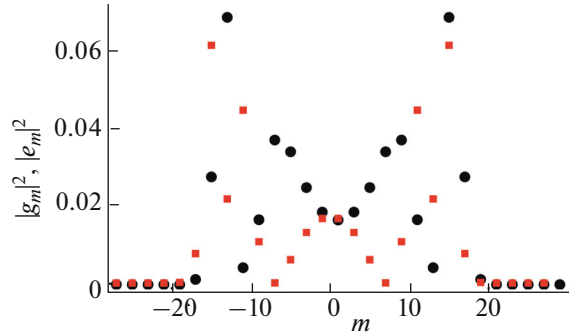

**Fig. 1.** Characteristic form of the probability distribution of momentum states for the ground (circles) and excited (squares) internal states of an atom generated by the standing wave field; $\tau_1 = \pi/100$ and $\zeta = 3 \times 10^2$.

Here we will adopt a weaker constraint on the interaction time: $\tau \ll 1$ [17]. Then the approximate solution for the amplitudes of the pulse states at the ground and excited levels will be represented in the form [17]

$$g_m(n,\tau) = f_n \frac{1+(-1)^m}{2} e^{i2(m+n)\tau} J_m\left(\frac{\zeta}{m+n}\sin(2(m+n)\tau)\right), \quad (3a)$$

$$e_m(n,\tau) = f_n \frac{1-(-1)^m}{2} e^{i2(m+n)\tau} J_m\left(\frac{\zeta}{m+n}\sin(2(m+n)\tau)\right), \quad (3b)$$

where it is assumed that before the interaction the atom is in the ground state, and $J_m(x)$ is the Bessel function. Note that the solution has a periodicity $T = 2\pi$. Finally, having accepted the fulfillment of the normalization condition $\sum_{m,n=-\infty}^{\infty}\left(|g_m(n,\tau)|^2 + |e_m(n,\tau)|^2\right) = 1$ as an indicator of the correctness of the approximation, one can easily verify that the applicability of formulas (4a) and (4b) is quite reliable within the limits $\zeta\tau \leq 10$.

Momentum distributions of an initially stationary sodium atom ($\omega_r \approx 1.5 \times 10^5$ Hz) after scattering on a standing wave are shown in Fig. 1 (at $\Omega_{\text{Rabi}} = dE/\hbar = 4.5 \times 10^7$ Hz, $t = 2 \times 10^{-7}$ s). As can be seen, the number of pulse states and the approximately identical nature of their distribution are quite sufficient to find the optimal situation for the subsequent implementation and registration of the interference of waves of atomic matter.

### 2.2. Formation of Asymmetry in the Momentum Distribution of an Atom

Immediately after the generation of the pulse states (5), the atom can be in free motion for some time from moment $\tau_1$ to moment $\tau_2 \geq \tau_1$. During this time, the coefficient wave functions of translational motion will be determined by the following expressions:

$$g(\eta,\tau) = \sum_{l=-\infty}^{\infty} g_l(\tau_1) e^{il\eta - il^2\tau}, \quad e(\eta,\tau) = \sum_{l=-\infty}^{\infty} e_l(\tau_1) e^{il\eta - il^2\tau}, \quad (4)$$

Next (from the moment $\tau_2$) the interaction of the atom continues with one of the counter waves creating a wave. The optical transitions induced by it superimpose in pairs the pulse states from the ground and excited levels of the atom, differing by one photon pulse, and thus carry out the interference of these matter waves [21]. As a result, the amplitudes of the pulse states in the wave functions written in the form

$$g(\eta,\tau) = \sum_{l=-\infty}^{\infty} g_l(\tau) e^{il\eta}, \quad e(\eta,\tau) = \sum_{l=-\infty}^{\infty} e_l(\tau) e^{il\eta}, \quad (5)$$

are represented by expressions

$$g_l(\tau) = c_{l,1} e^{-i\lambda_{l,1}(\tau-\tau_2)} + c_{l,2} e^{-i\lambda_{l,2}(\tau-\tau_2)}, \quad e_l(\tau) = d_{l,1} e^{-i\mu_{l,1}(\tau-\tau_2)} + d_{l,2} e^{-i\mu_{l,2}(\tau-\tau_2)}, \quad (6)$$

where

$$c_{l,1} = -\frac{\varsigma^* e_{l+1}(\tau_2) + (\lambda_{l,2} - l^2) g_l(\tau_2)}{\lambda_{l,1} - \lambda_{l,2}}, \quad c_{l,2} = \frac{\varsigma^* e_{l+1}(\tau_2) + (\lambda_{l,1} - l^2) g_l(\tau_2)}{\lambda_{l,1} - \lambda_{l,2}},$$

$$d_{l,1} = -\frac{\varsigma g_{l-1}(\tau_2) + (\mu_{l,2} - l^2) e_l(\tau_2)}{\mu_{l,1} - \mu_{l,2}}, \quad d_{l,2} = \frac{\varsigma g_{l-1}(\tau_2) + (\mu_{l,1} - l^2) e_l(\tau_2)}{\mu_{l,1} - \mu_{l,2}},$$

$$\lambda_{l,1(2)} = \frac{1}{2} + l + l^2 \mp \sqrt{\frac{1}{4} + l + l^2 + |\varsigma|^2}, \quad \mu_{l,1(2)} = \frac{1}{2} - l + l^2 \mp \sqrt{\frac{1}{4} + l + l^2 + |\varsigma|^2}, \quad (7)$$

$$g_l(\tau_2) = e^{i 2 l \tau_1 - i l^2 \tau_2} \sum_{s=-\infty}^{\infty} i^{\frac{l-s}{2}} f_{\frac{l+s}{2}} \frac{1 + (-1)^{\frac{l-s}{2}}}{2} J_{\frac{l-s}{2}}\left(\frac{\varsigma}{l} \sin(2l\tau_1)\right),$$

$$e_l(\tau_2) = e^{i 2 l \tau_1 - i l^2 \tau_2} \sum_{s=-\infty}^{\infty} i^{\frac{l-s}{2}} f_{\frac{l+s}{2}} \frac{1 - (-1)^{\frac{l-s}{2}}}{2} J_{\frac{l-s}{2}}\left(\frac{\varsigma}{l} \sin(2l\tau_1)\right).$$

Formulas (5) and (6) complete the description of the generation and interference of matter waves, which are subject to further measurement at the output of the atomic interferometer.

It is to be that the momentum distribution at the output of the interferometer contains two types of interference. One of them is inherent in the problem of scattering on a periodic potential and is represented by the sum of the momentum states in the expression of the wave function (5). The second is the key dynamic process in the atomic interferometer proposed here. It superimposes the corresponding momentum states of the ground and excited energy levels of the atom and is represented by two sums of two terms in the numerators (7).

## 3. REGISTRATION OF INTERFERENCE BY POLARIMETRIC METHOD

To measure the state (7), entangled between the translational and internal degrees of freedom of the atom, a polarimetric method is proposed in which the pump field (standing wave and subsequent traveling wave) is circularly polarized and couples the magnetic sublevels of a two-level atom, as shown in Fig. 2. It approximately equalizes the populations of the two sublevels, simultaneously forming on each of them discrete set of impulse states containing the required information about the interference of waves of atomic matter. In this case, the atomic medium (weakly interacting atoms) becomes optically gyrotropic.

Propagating along the pump line, the probe field of linear polarization is represented as the sum of two oppositely rotating circular polarizations. The probe field component with circular polarization of the pump field activates the optical transition $j_{1,z} = -1/2 \leftrightarrow j_{2,z} = 1/2$, and reverse polarization activates the optical transition $j_{1,z} = 1/2 \leftrightarrow j_{2,z} = -1/2$, not perturbed by the pump field (Fig. 2). Therefore, the two circular components of the probe wave in the atomic medium propagate with different phase velocities, as a result of which the total linear polarization at the output rotates relative to the input direction by a certain angle. In this case, the momentum distribution of the atom - the carrier of information about the interference of matter waves — is uniquely displayed because of the Doppler effect on the frequency spectrum of the rotated component of the probe wave.

The mathematics for before interaction with the pumping field computing the rotating component of the probing field is well known [16] and this situation is given

$$E_{p,y}(z,t) = \frac{E_p(0,t)}{2}\left[\exp(-iqF(\omega_p)k_p z) - \exp\left(-i\frac{q}{\Delta_p - k_p v_r/2 + i\gamma} k_p z\right)\right], \quad (8)$$

where $q = \pi N |d|^2 / 6\hbar$, $N$ is the atomic density at the ground state, uniformly distributed between the magnetic sublevels $j_{1,z} = \pm 1/2$ before interaction with the pumping field, $k_p = \omega_p/c$, $\omega_p$ is the circular frequency of the probing wave, $d$ is the reduced matrix element of optical transition, $\Delta_p = \omega_p - \omega_0$, $v_r = \hbar k/M$ is the single-photon recoil rate, and

$$F(\omega_p) = \sum_{l=-\infty}^{\infty} \frac{|g_l(\tau_3)|^2}{\Delta_p - 2(l+1/2)k_p v_r + i\gamma},$$

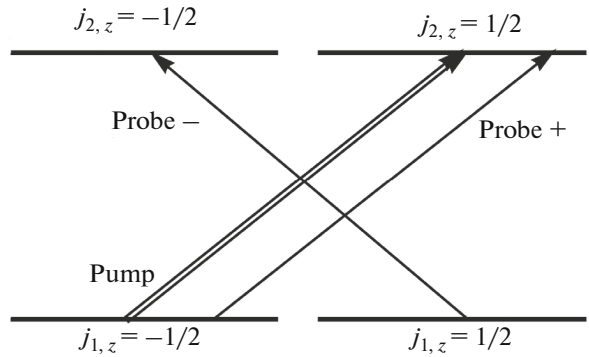

**Fig. 2.** A circularly polarized pump field induces optical gyrotropy in the atomic medium (a thin layer of laser-cooled atoms), rotating the linear polarization of the probing radiation.

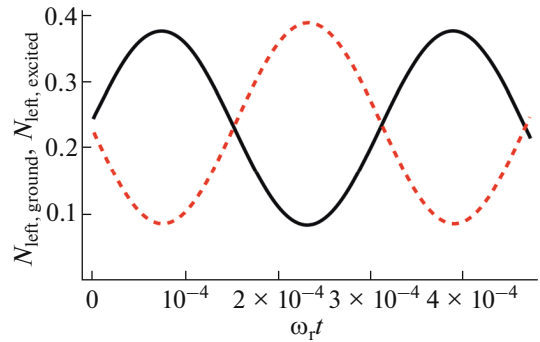

**Fig. 3.** Time evolution of negative-momentum states of an atom at the excited (dashed line) and ground (solid line) levels of internal energy; $\tau_2 = \tau_1 = \pi \times 10^{-3}$ and $\varsigma = 10^4$.

where $t_3$ is the moment of switching off the traveling pump wave. The parameter of inhomogeneous broadening of optical lines $\gamma$ is introduced phenomenologically.

For practical applications of the method, it is desirable to optimize and, if possible, simplify the nature of the output signal (8). In many ways, this goal can be achieved by choosing the parameters of the traveling pump wave, which swings the distributions of pulses at the ground and excited levels of the atom in opposite directions with a large amplitude. This is a swing for the population of the left part of the distribution (i.e. for the region $m \leq 0$) on the ground and excited states are illustrated in Fig. 3 [21], creating a kind of Schrödinger cat in momentum space. In our opinion, the most optimal moments are when the distributions of pulses on the ground and excited magnetic sublevels are maximally distant from each other (the minimum and maximum points in Fig. 3). Polarization spectroscopy is left with the task of projecting these highly asymmetric momentum distributions into the spectral distribution of the rotating component of the probe wave using the Doppler effect. Figure 4 shows the result of such a projection for the conditions of the first extremum of Fig. 3.

Suppose some preliminary information about the system is needed. In that case, measuring the spectrum directly behind the counter-propagating waves, while it is still symmetrical about the initial (zero) value, is convenient.

## 4. CONCLUSION

A new type of atomic interferometer is proposed, combining the well-developed areas of atomic optics and polarization spectroscopy. The first stage of interaction consists of generating equidistant pulse states because of the diffraction of a two-level atom on a resonant standing electromagnetic wave. Further interference is caused by momenta at the ground and excited internal energy levels of the atom in opposite directions. The measuring stage is based on polarization spectroscopy, in which the interference result of atomic momentum states is reproduced in the spectral distribution of the rotated component of the probing electromagnetic radiation. Optimal conditions for outputting information about the interference process are determined, and it is shown that the usual conditions of specialized laboratories for laser cooling of alkali metal vapors (for example, those selected in Fig. 4) are more than sufficient for the successful operation of the proposed atomic interferometer.

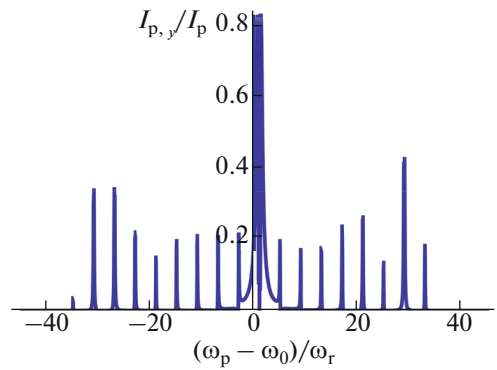

**Fig. 4.** Spectrum of the rotated polarization of the probing field at the moment of the first maximum of the asymmetry of the momentum distribution (Fig. 3). The thickness of the atomic sample is 1 μm, and the density is $10^{11}$ cm$^{-3}$. These values are typical for magneto-optical traps; $\tau_2 = \tau_1 = \pi \times 10^{-3}$ and $\varsigma = 3 \times 10^2$.

It should also be added that because of the oscillatory nature of the pulse distribution, the average spatial displacements of the atoms are less than the wavelength of the optical fields, i.e., the trajectories of atoms of different pulse states remain overlapped all the time. This indicates the compactness of the polarimetric atomic interferometer design, as well as the prospects for its portable version.

## FUNDING

The work was carried out with the financial support of the Committee on Higher Education and Science of the Republic of Armenia within the framework of the Laboratory of Research and Modeling of Quantum Phenomena of the Institute of Physics of Yerevan State University.

## CONFLICT OF INTEREST

The author of this work declares that he has no conflicts of interest.

*Translated by V. Musakhanyan*

**Publisher's Note.** Pleiades Publishing remains neutral with regard to jurisdictional claims in published maps and institutional affiliations.
AI tools may have been used in the translation or editing of this article.